\documentclass[jcp,aip,reprint]{revtex4-1}
\usepackage{bm}
\usepackage{epsfig}

\usepackage{amsmath}

\newcommand{\ee}{\mathrm{e}}

\newcommand{\dd}{\mathrm{d}}

\newcommand{\CC}{\mathcal{C}}

\newcommand{\tobs}{t_\mathrm{obs}}
\newcommand{\salt}{s_\mathrm{alt}}
\newcommand{\Kalt}{K_\mathrm{alt}}
\newcommand{\kalt}{k_\mathrm{alt}}
\newcommand{\Veff}{V_\mathrm{eff}}

\newcommand{\etal}{\emph{et~al.}}

\begin{document}

\title{Dynamical phase transitions in supercooled liquids: interpreting measurements of dynamical activity}
\author{Christopher J. Fullerton} 
\affiliation{Department of Physics, University of Bath, Bath, BA2 7AY}
\author{Robert L. Jack}
\affiliation{Department of Physics, University of Bath, Bath, BA2 7AY}

\begin{abstract}
We study dynamical phase transitions in a model supercooled liquid. These transitions occur
in ensembles of trajectories that are biased towards low (or high) dynamical activity.  We compare
two different measures of activity that were introduced in recent papers and we find that they
are anti-correlated with each other.  To interpret this result, we show that the two measures
couple to motion on different length and time scales.  We find that `inactive' states with very slow
structural relaxation nevertheless have increased molecular motion on short scales.  We discuss these
results in terms of the potential energy landscape of the system and in terms of the liquid structure in active/inactive
states.
\end{abstract}

\maketitle

\section{Introduction}
\label{sec:intro}

As liquids are cooled towards their glass transitions, their relaxation times increase dramatically, and
the motion of their constituent particles becomes increasingly co-operative and heterogeneous~\cite{ediger96,ediger00,deb-still01}.
There are several competing theories that aim to describe these phenomena~\cite{chandler10-annrev,ktw89,gotze92,still95},
but neither simulation nor experimental data have so far proven sufficient to establish which (if any)
can fully describe the supercooled liquid state.  Recently, novel dynamical phase transitions 
have been discovered in glassy systems~\cite{garrahan-fred07,garrahan-fred09,jack10-rom,elmatad10-pnas}:
these are new results that can be used to test existing theories.
These phase transitions take place in ensembles of trajectories (sometimes called $s$-ensembles), 
where the dynamical evolution
of the glassy systems is biased towards low-activity states~\cite{merolle05,jack06-spacetime,lecomte07,garrahan-fred09}.  
Since these phase transitions are dynamical
in nature, they fit naturally with theories of the glass transition where dynamical motion takes a central
role~\cite{gc02-prl,gc03-pnas,chandler10-annrev}, but they can also be interpreted in terms of random first order transition theory~\cite{ktw89}, and 
are linked with properties of the energy landscape and its normal modes~\cite{still95,coslovich06,manning11,reichman08-modes}. 

In this article, we discuss these dynamical phase transitions and their associated ensembles of trajectories.
We are motivated primarily by two previous studies~\cite{hedges09,pitard11} which provided evidence for such transitions
in a model glass-former, composed of Lennard-Jones particles~\cite{ka95a,ka95b}.
In the first study, Hedges~\etal~\cite{hedges09} measured the activity in this model through the mean square
displacement of its particles.  Biasing the dynamics with respect to this parameter,
they found evidence for 
a first-order phase transition between active (equilibrium fluid) and inactive (glass) states.
In the second study, Pitard~\etal~\cite{pitard11} used an alternative measure of activity, based on the steepness
and curvature of the energy landscape, integrated over time.  Using this activity measure to bias the
system, they again found evidence for a dynamical phase transition, but the properties of the dynamical phases
were different to those found in Ref.~\onlinecite{hedges09}, including apparently non-extensive behaviour of the activity in
one of the phases.

In this study, we combine measurements of the different measures of activity used in Ref.~\onlinecite{hedges09,pitard11}.  We find that
these measures couple to different kinds of molecular motion.  Further, the two measures are 
anti-correlated in the system that we consider.  Physically, this happens because stable states with
very slow structural relaxation may have an increased propensity for `vibrational' motion (or $\beta$-relaxation)
on short length scales. Based on this observation, we are able to resolve some of the apparent differences
between the results of Ref.~\onlinecite{hedges09,pitard11}.
We also gain insight into the nature of the inactive (glassy) states, and how these relate
to properties of the underlying energy landscape, and the normal modes associated with motion on this
landscape.

Section~\ref{sec:background} of this paper introduces the model and the ensembles that we will use; in 
Sec.~\ref{sec:measurements}, we 
compare the two measures of the activity used in Ref.~\onlinecite{hedges09,pitard11}, showing that they
are anti-correlated.  In Sec.~\ref{sec:interpret}, we investigate the activity of Pitard~\etal~\cite{pitard11}
in more detail, and discuss the relationship
of this activity measurement to other properties of the fluid and glassy states in the system.  
We summarise our main conclusions in Sec.~\ref{sec:conc}.

\section{Background}
\label{sec:background}

\subsection{Model}

We consider the Kob-Andersen mixture of Lennard-Jones particles~\cite{ka95a,ka95b}, which is a well-studied
model glass former.
There are $N$ particles in the system and a configuration $\bm{r}^N$ has potential energy $E(\bm{r}^N) = \sum_{i<j} V(r_{ij})$,
where $r_{ij}$ is the distance between particles $i$ and $j$, and
\begin{align}
 V_{ij}(r_{ij}) = 4\epsilon_{ij} \left[ \left( \frac{\sigma_{ij}}{r_{ij}} \right)^{12} - \left(\frac{\sigma_{ij}}{r_{ij}}\right)^6\right].
\end{align}
There are two species of particle, A (large) and B (small) and the parameters $\epsilon_{ij}$ and $\sigma_{ij}$ depend on the species of particles $i$ and $j$, as 
$\sigma_{\mathrm{AA}} = \sigma = 1$, $\sigma_{\mathrm{BB}} = 0.88\sigma$, $\sigma_{\mathrm{AB}} = 0.8\sigma$, $\epsilon_{\mathrm{AA}} = \epsilon = 1$, $\epsilon_{\mathrm{BB}} = 0.5\epsilon$ and $\epsilon_{\mathrm{AB}} = 1.5\epsilon$.
For numerical efficiency, $V_{ij}(r_{ij})$ is truncated at $r_{ij}^{\mathrm{cut}} = 2.5\sigma_{ij}$ and shifted so that the energy of a pair of particles separated by $r_{ij}^{\mathrm{cut}}$ is zero.
For a system of $N$ particles, there are $N_{\rm A}=(4N/5)$ particles of type A and $N_{\rm B}=(N/5)$ of type B.
The density is fixed at $\rho = 1.2\sigma^{-3}$ as in Ref.~\onlinecite{hedges09}: note that $\rho=1000/(9.4\sigma)^3\approx1.204\sigma^{-3}$ was used
in Ref.~\onlinecite{ka95a,ka95b} and in some other studies.  This small difference has no qualitative effect on the behaviour shown here.

The system evolves by Monte Carlo (MC) dynamics: as discussed by Berthier and Kob~\cite{berthier-kob07}, this dynamical scheme
results in structural relaxation that is in quantitative agreement with molecular dynamics, up to a rescaling of time.  It was also
shown in Ref.~\onlinecite{hedges09} that MC dynamics and constant-temperature molecular dynamics gave very similar results in the $s$-ensemble.
The MC dynamical scheme corresponds to a system evolving with
overdamped Langevin dynamics,
\begin{align}
\frac{\partial \bm{r}_i}{\partial t} = -\beta \nabla_i E + \bm{\eta}_i(t),
\label{eq:langevin}
\end{align}
where $D_0$ is the (bare) diffusion constant of a single particle, $\beta=1/T$ is the inverse temperature (we take Boltzmann's constant $k_\mathrm{B}=1$),
and $\bm{\eta}_i(t)$ is white noise with zero mean, and covariances
\begin{align}
\langle \eta^{\mu}_i(t) \eta^{\nu}_j(t') \rangle = 2 D_0 \delta_{ij} \delta^{\mu \nu} \delta(t-t'),
\end{align}
in which $\mu$ and $\nu$ label cartesian components of the vector $\bm{\eta}(t)$.
The natural units in the system are the length $\sigma$ (the diameter of a large particle); the energy $\epsilon$ (interaction strength between large particles); and the time scale $\Delta t = \sigma^2/D_0$ (of the order of the Brownian time for a free particle).
When discussing our numerical results in the following sections, 
we take $(\sigma,\epsilon,\Delta t)$ all equal to unity, for compactness.

The MC dynamical scheme that we use is equivalent to the Langevin equation (\ref{eq:langevin}) in the limit when all MC steps are small (see for example Ref.~\onlinecite{whitelam11-molsim}).
As in Ref.~\onlinecite{berthier-kob07}, we draw trial MC displacements from a cube of side
$\delta = 0.15 \sigma$, centred on the origin. This choice of step size leads to efficient simulations which accurately capture the
nature of the structural relaxation.
The mean square displacement for a trial MC move is $\delta^2/4$: the requirement that the diffusion constant be $D_0=\sigma^2/\Delta t$ means
that $\Delta t$ corresponds to $24(\sigma/\delta)^2 \approx 1070$ MC sweeps.

We emphasise that overdamped dynamics as studied here were used by Hedges~\etal~\cite{hedges09}, who also considered molecular
dynamics with a strong coupling to a thermostat.  However, the results of Pitard~\etal~\cite{pitard11} were obtained using
molecular dynamics at constant energy.

\subsection{Ensembles of trajectories, and measures of activity }

\newcommand{\rnt}{\bm{r}^N\!(t)}

We consider dynamical transitions that occur in ensembles of trajectories.  These trajectories
have duration $\tobs$, and each trajectory is divided into $M$ ``slices'', each of duration $\Delta t$.
Following Hedges~\etal~\cite{hedges09}, the activity of a trajectory $\rnt$ is defined as
\begin{align}
\displaystyle K[\rnt] = \Delta t \sum_{i=1}^{N_{\rm A}} \sum_{j=0}^M | \bm{r}_i(t_j) - \bm{r}_i(t_{j-1}) |^2
\label{eq:def-K}
\end{align}
where the index $i$ runs over all particles of type A, and the $t_j$ are the times that separate the slices: $t_j = j\Delta t$.
We also define the intensive ``activity density'' $k = K/(N_{\mathrm{A}}\tobs)$, which we sometimes refer to simply as the activity.

From (\ref{eq:def-K}), it follows that $k$ measures the mean square displacement of a type-A particle during a time interval $\Delta t$.
This time scale is comparable with the time taken for a free particle to diffuse over its own diameter; in the supercooled
state then $\Delta t$ is long enough for a particle to explore its local environment (part of the $\beta$-relaxation process),
but $\Delta t$ is shorter than the typical time for the fluid structure to relax (the $\alpha$-process).
Our interpretation is that $k$ measures motion on length scales comparable to the particle diameter.

The dynamical phase transitions that we will consider occur when the equilibrium ensemble of trajectories is biased to
low activity.  We define a biased ensemble (or `$s$-ensemble') through its probability distribution over trajectories:
\begin{align}
P_s[\rnt] \propto  P_0[\rnt]
\ee^{-s K[\rnt]}
,
\label{eq:traj_prob}
\end{align}
where $P_0[\rnt]$ is the equilibrium probability of trajectory $\bm{r}^N(t)$.  (In defining
the probability distributions over trajectories, it is sufficient for our purposes to represent a trajectory
as the set of $M+1$ configurations at the times $t_j$ that separate the slices.  However, a finer-grained representation
in time is also possible.)

Within the $s$-ensemble the average of any trajectory-dependent observable $A$ may calculated using
\begin{align}
\langle A \rangle_s = \frac{\langle A \ee^{-sK} \rangle_0}{\langle \ee^{-sK} \rangle_0},
\label{eq:ave-s}
\end{align}
where $\langle \cdot \rangle_s$ denotes an average over trajectories of length $\tobs$ in the $s$-ensemble and $\langle \cdot \rangle_0$ means an average of trajectories of length $\tobs$ at equilibrium (which corresponds to $s=0$).

An alternative measure of the activity was proposed by Pitard~\etal~\cite{pitard11}, as the time integral (between $t=0$ and $t=\tobs$) of
an `effective potential':
\begin{align}
\displaystyle \Veff = \frac{\beta}{4} \sum_i |\bm{F}_i|^2 + \frac12 \sum_i \nabla_i\cdot\bm{F}_i,
\label{eq:def-veff}
\end{align}
where the index $i$ runs over all particles and $\bm{F}_i = -\nabla_i E$ is the force on particle $i$.

In this study, we define 
\begin{align} 
\Kalt[\rnt] = \frac{\Delta t}{2} \sum_{j=1}^M \left[ \Veff(t_{j-1}) + \Veff(t_{j}) \right]
\label{eq:kaltdefine}
\end{align}
which is an estimate of the integral of $\Veff$, using a trapezium rule (we take $t_j=j\Delta t$ as above).  
The notation $\Kalt$ indicates that this is an `alternative' activity.  
We also define $\kalt = \Kalt/(N\tobs)$, by analogy with $k$.
Since $\Veff$ is evaluated at only $M+1$ points
within the trajectory, $\Kalt$ is not a very precise estimate of the integral of $\Veff$ proposed by Pitard~\etal~\cite{pitard11}
as an activity measure.  
However, we expect that $\Kalt$ captures the same physical features as this measurement.
We also performed simulations where $2M + 1$ points were used to calculate $\Kalt$ (the step size in the trapezium rule was halved).
This produced no qualitative difference in the values of $\Kalt$ we obtained. This means that $M + 1$ points are sufficient to make $\Kalt$ a good estimate of the integral of $\Veff$.

The relation between $\Kalt$ and dynamical activity is not obvious \emph{a priori}.  Pitard~\etal~\cite{pitard11} identified $\Kalt$ as
an activity by considering
the probability that a particle returns to (or remains at) its original position over a small time $\delta t$.
This probability is obtained from the propagator $G(\bm{r}'^N,t';\bm{r}^N,t)$ which gives the probability that a system in configuration
$\bm{r^N}$ at time $t$ will evolve into configuration $\bm{r}'^N$ at time $t'$.  
For Langevin dynamics as considered here, the probability that the initial and final states are the same
is given by
Autieri~\etal~\cite{autieri09}: for small $\delta t$,
\begin{align} 
G(\bm{r}^N,t+\delta t; \bm{r}^N,t) & = z^{-1} \frac{\ee^{-\beta \Veff \delta t+{\cal O}(\delta t^2)}}{ (\delta t)^{3N/2}}
\nonumber\\ &
 = z^{-1} \ee^{-(3N/2)\log \delta t - \beta \Veff \delta t+{\cal O}(\delta t^2)},
\label{eq:propagator}
\end{align}
where $z$ is a normalisation constant (independent of time).  We include the full dependence of $G$ on $\delta t$ to emphasise
that $G$ decreases with $\delta t$, regardless of the sign of $\Veff$.  (On setting $\Veff=0$, one recovers the standard result
for $N$ non-interacting Brownian particles.)
Equ.~(\ref{eq:propagator}) shows that when $\Veff$ is large then particles in the system are likely to move quickly away from their
original positions; when $\Veff$ is small then particle are more likely to remain localised.  This is the motivation for 
proposing $\Kalt$ as a measure of dynamical activity.  Note however that this measurement is defined in terms of motion
on the very small time scale $\delta t$.

Following Pitard~\etal~\cite{pitard11}, we therefore define an `$\salt$-ensemble' through a bias on $\Kalt$:
\begin{align}
P_{\salt}[\rnt] \propto P_0[\rnt] \ee^{-\salt \Kalt[\rnt]}.
\label{eq:traj_prob_salt}
\end{align}
This definition is analogous to (\ref{eq:traj_prob}): continuing the analogy for averages of
an observable $A$, we have
\begin{align}
\langle A \rangle_{\salt} = \frac{\langle A \ee^{-\salt\Kalt } \rangle_0}{\langle \ee^{-\salt\Kalt} \rangle_0},
\end{align}
by analogy with (\ref{eq:ave-s}).  Equations (\ref{eq:traj_prob}) and (\ref{eq:traj_prob_salt}) define the ensembles
of trajectories that we will consider in the following.

\section{Measurements of activities in biased ensembles}
\label{sec:measurements}

\begin{figure}
\includegraphics[width=8.5cm]{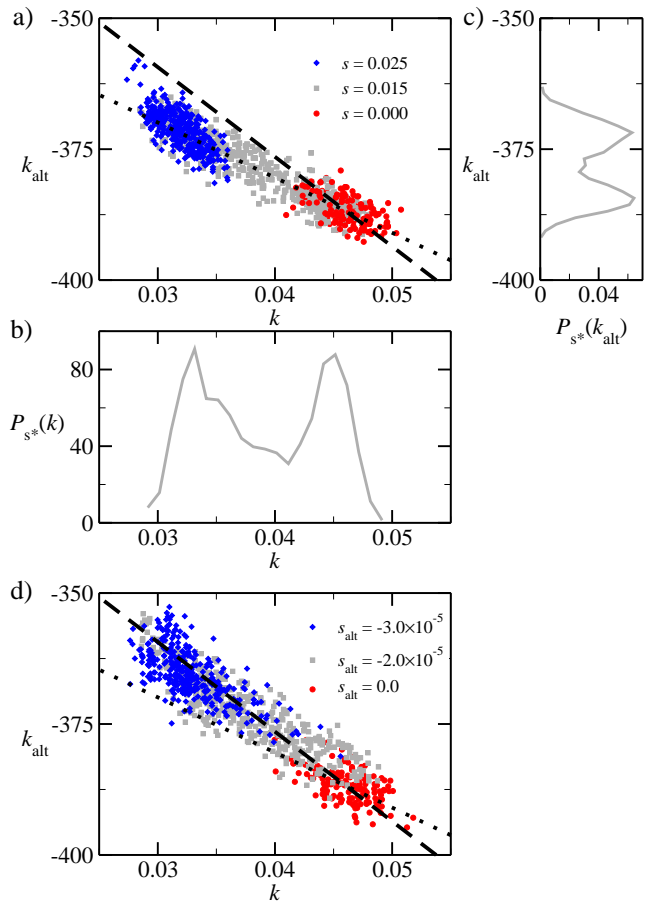}
\caption{
(a)~Scatter plot of the two activity measurements $k$ and $\kalt$, in three different $s$-ensembles.  
The ensembles are characteristic of the active phase ($s = 0.000$), the coexistence region ($s = 0.015$) and the inactive phase ($s = 0.025$).
The two activity measurements $k$ and $\kalt$ are anti-correlated.  The trajectory length is $\tobs=400\Delta t$. 
(b, c)~Marginal distributions of $k$ and $\kalt$ from the $s$-ensemble with $s=0.015$. 
This bimodal behaviour is characteristic of the dynamical phase transition found in Ref.~\onlinecite{hedges09}.
(d)~Scatter plot of $k$ and $\kalt$ for three values of $\salt$ and $\tobs=200\Delta t$. The data for
 $\salt = -3.0\times 10^{-5}$ is similar to the inactive data for $s=0.025$.
The dashed and dotted lines in (a) and (d) are the same in both panels and are obtained by linear regression
analyses on data from (a) for the dots and (d) for the dashes.
}
\label{fig:scatter}
\end{figure}

\begin{figure}
\includegraphics[width=7.5cm]{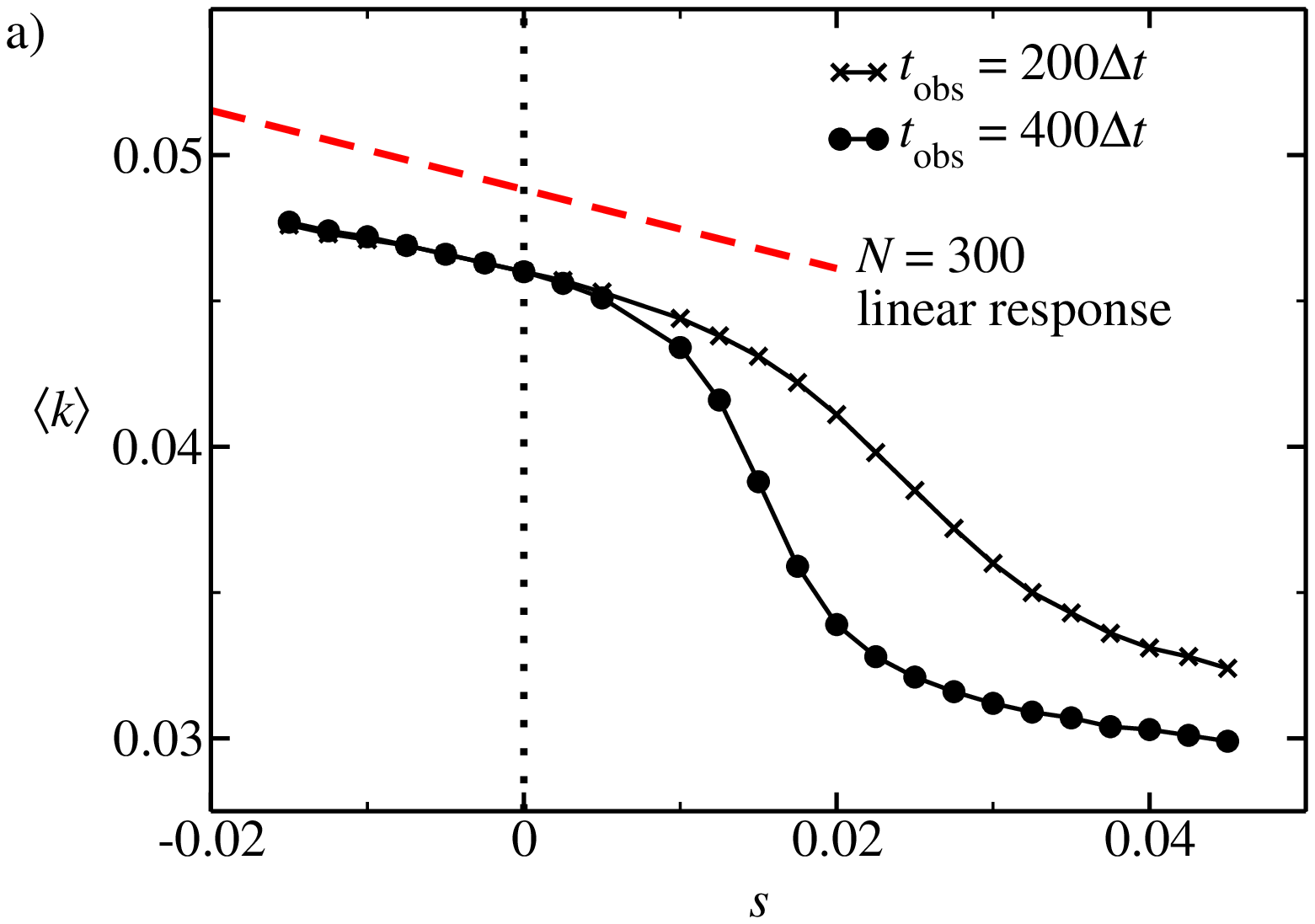}
\includegraphics[width=7.5cm]{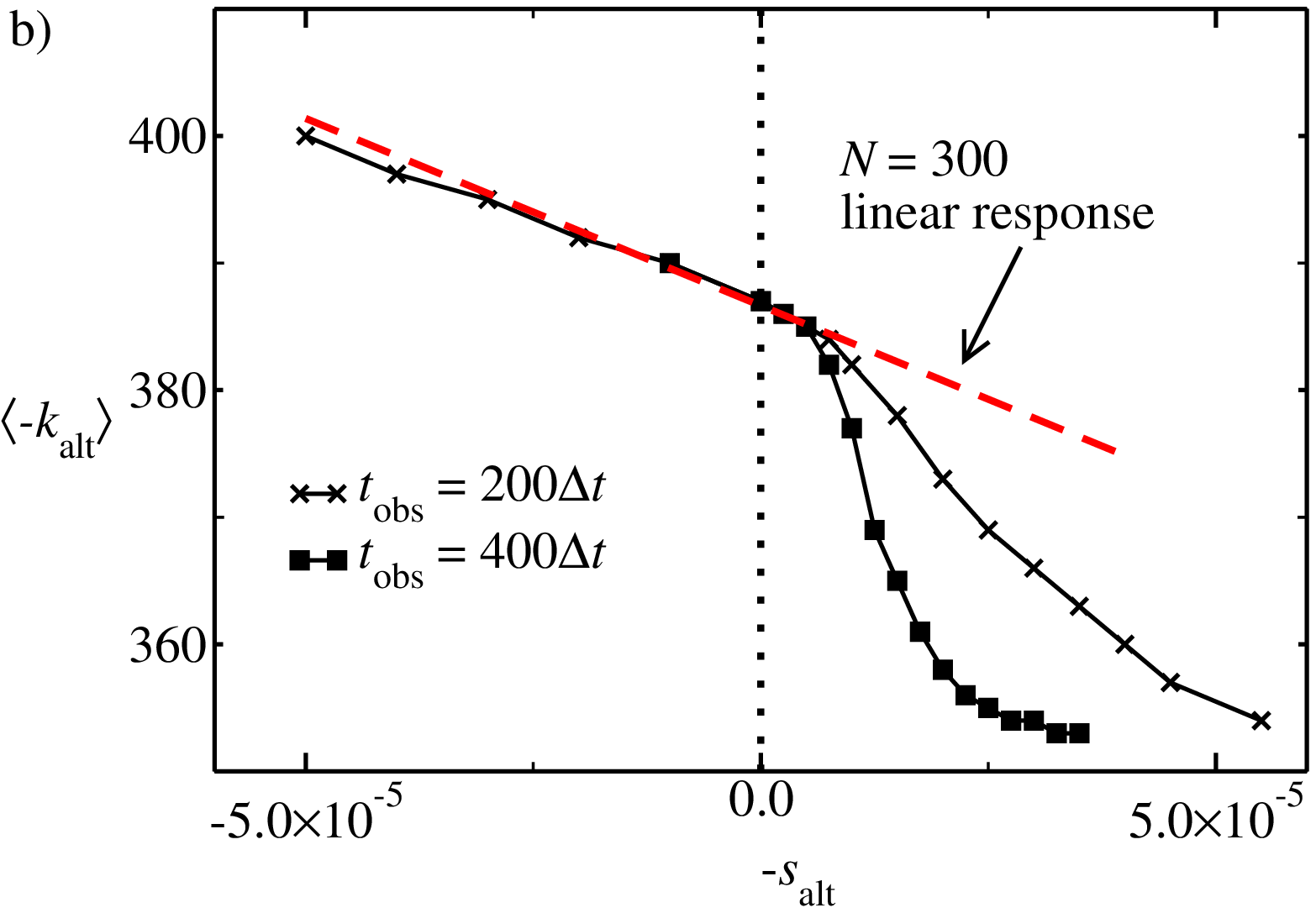}
\caption{Averaged activities in biased ensembles.  
Note that panel (b) shows the negatives of the field and the activity, $-\salt$ and $\langle-\kalt\rangle$.
All data are for $N=150$ and $T=0.6$, except
for the red dashed lines, where $N=300$ and we show the linear response behaviour about equilibrium: 
$\langle K \rangle_s = \langle K \rangle_0 + s \langle \delta K^2 \rangle_0 + O(s^2)$,
and similarly for $\salt$.  These linear response results do not capture the non-trivial crossovers, but they do show that the mean
and variance of $K$ and $\Kalt$ are approximately extensive in $N$, for $s=0$ (there is a weak finite-size correction to $\langle k\rangle_0$:
particle motion in smaller systems is known to be slightly slower for this system, compared to the bulk~\cite{berthier12-fsize}).
}
\label{fig:K_alt_K_s}
\end{figure}

We use transition path sampling (TPS)~\cite{tps-annrev02} to sample biased ensembles of trajectories, as discussed in Appendix~\ref{app:sampling}.
We show numerical results obtained by TPS in Figs.~\ref{fig:scatter} and~\ref{fig:K_alt_K_s},
which summarise the behaviour of $K$ and $\Kalt$, as $s$
and $\salt$ are varied.  We concentrate on the behaviour of a system of $N=150$ particles
at temperature $T=0.6$, as in Ref.~\onlinecite{hedges09}. 
[Recall we have fixed units such that $(\epsilon,\sigma,\Delta t)$ are all equal to unity.]
In Figs.~\ref{fig:scatter}(a,d), we show scatter plots of $K$ and $\Kalt$, combining data sampled from equilibrium and
for several values of $s$ and $\salt$.
We find that $k$ is always positive and $\kalt$ is always negative. (It is worth noting that throughout this article
we write ``$\kalt$ is larger than $\kalt'$'' if $\kalt>\kalt'$, regardless of the sign of $\kalt$.)  Perhaps surprisingly,
we also find that while $k$ and $\kalt$ were both proposed as measures of dynamical
activity, they are anti-correlated with one another.  This observation will be crucial in the following discussion.

Panels (b) and (c) of Fig.~\ref{fig:scatter} also show that for an appropriate value of $s$ (here $s=s^*=0.015$),
the marginal distributions of both $k$ and $\kalt$ are bimodal.  These distributions are indicative of a dynamical
phase transition, although the existence of such a transition can be confirmed only if these distributions
remain bimodal as the system size $N$ and observation time $\tobs$ are taken to infinity.  

In Fig.~\ref{fig:K_alt_K_s}, we show average values of $k$ and $\kalt$ in ensembles of trajectories, as $s$ and
$\salt$ are varied.  We note however that in Fig.~\ref{fig:K_alt_K_s}(b), we are plotting $\langle -\kalt \rangle$ against
$-\salt$.  On increasing $s$ in panel (a), we observe a crossover from a large-$k$ state at $s=0$ to a small-$k$ state
at positive $s$.  As in Ref.~\onlinecite{hedges09}, this crossover becomes sharper as $\tobs$ is increased, consistent with a dynamical 
first-order phase transition.  As we increase $-\salt$ (or decrease $\salt$) in Fig.~\ref{fig:K_alt_K_s}(b), we observe a similar crossover to a state
with smaller $-\kalt$ (and hence larger $\kalt$).  Again, the crossover sharpens on increasing $\tobs$.

Finally, returning to Fig.~\ref{fig:scatter}(a,d), we observe that the states for $s=0.025$ and $\salt=-3\times10^{-5}$ have
similar joint distributions of $(k,\kalt)$.  Hence, taking Figs.~\ref{fig:scatter} and~\ref{fig:K_alt_K_s} together, we 
infer that the two crossovers shown in Fig.~\ref{fig:K_alt_K_s} represent
transitions between the same two states: the equilibrium state [colored red in Fig.~\ref{fig:scatter}(a,d)] and the state
that was identified by Hedges~\etal~as the glassy (inactive) state [colored blue in Fig.~\ref{fig:scatter}(a,d)].
It was shown by Hedges~\etal~\cite{hedges09} that the inactive state was accompanied by a self-intermediate scattering function
that does not decay throughout the observation time $\tobs$, indicating that particles remain localised near their
initial positions throughout the trajectory.  Our data confirm this result: this is the sense in which this small-$k$ state is `inactive'.

The crossover shown in Fig.~\ref{fig:K_alt_K_s}(b) for small negative $\salt$ was not reported by Pitard~\etal~\cite{pitard11}.  However, we note that the
ranges of $\salt$ and $\Kalt$ shown in Fig.~\ref{fig:K_alt_K_s}(b) are much smaller than those used in Ref.~\onlinecite{pitard11}.  It is possible
that a more detailed analysis of the relevant range of $\salt$ using the methodology of Ref.~\onlinecite{pitard11} 
might reveal a similar crossover/transition.
What is clear from Figs.~\ref{fig:scatter} and~\ref{fig:K_alt_K_s} is that the transition (for $\salt>0$)
reported by Pitard et al~\cite{pitard11} is a different phenomenon to that reported by Hedges et al~\cite{hedges09}.

The transition reported by Pitard et al~\cite{pitard11} for $\salt>0$ is accompanied by anomalous behaviour of
the derivative $\mathrm{d}\kalt/\mathrm{d}\salt$ and non-extensivity of $\kalt$ itself, 
for small positive $\salt$ (and perhaps even for $\salt=0$).
For the narrow range of $\salt$ that we considered, we did not observe these effects. Fig.~\ref{fig:K_alt_K_s}(b) shows
that $\mathrm{d}\kalt/\mathrm{d}\salt$ depends very weakly on $\salt$ for $\salt<0$, and that on increasing the system
size to 300 particles, there is no significant in change either the equilibrium average of $\kalt$ nor in its derivative with respect 
to $\salt$.  The differences between our results and those of Ref.~\onlinecite{pitard11} in this regime remain a subject for future study: here
we concentrate on the crossover that we do find for $\salt<0$, and its relationship to the active/inactive phase coexistence
phenomena found in Ref.~\onlinecite{hedges09}.

\section{Interpretation of activity measurements}
\label{sec:interpret}

The interpretation of the activity $k$ is transparent in that it measures particle motion on a timescale
$\Delta t$.  As discussed in Ref.~\onlinecite{hedges09}, the low-$k$ phase found on increasing $s$ is characterised by an absence
of structural relaxation (at least for small systems of 150 particles, on time scales up to $40$ times the equilibrium
relaxation time).  The relation between $\kalt$ and particle motion is somewhat indirect, operating via the
expression (\ref{eq:propagator}) which gives the probability that a particle deviates significantly from its initial
position, on short time scales.

In the following, we focus on the activity $\kalt$, aiming in particular to understand why this activity measurement is larger
in the `inactive state' of Ref.~\onlinecite{hedges09}, compared with equilibrium.  

\subsection{Two contributions to $\Veff$, and a quasi-equilibrium/two-temperature scenario}

\begin{figure}
\includegraphics[width=7.5cm]{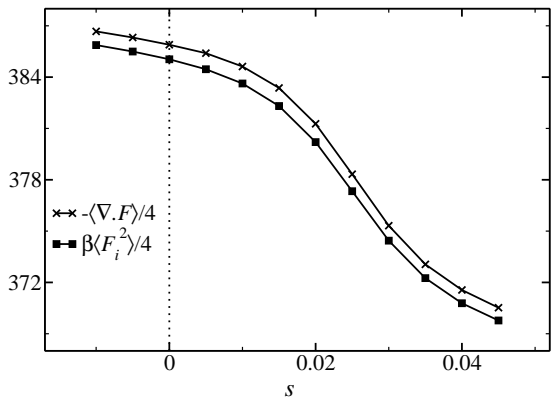}
\caption{
Numerical test of equation (\ref{eq:F_div_F}). The two quantities should be equal at equilibrium ($s=0$), but there is a small
difference between them due to our use of a truncated potential (see Appendix \ref{eq:F_div_F}). The small difference
is almost constant for the range of $s$ considered.
}\label{fig:FdivF}
\end{figure}

From (\ref{eq:def-veff}), we see that $\Veff$ (and hence also $\kalt$) has two contributions, one from the interparticle forces
and the other from the divergence of the force.
At equilibrium, these contributions are related:
\begin{align}
\begin{split}
\langle |\beta \bm{F}_i|^2 \rangle_0 =& Z^{-1} \int\dd\bm{r}^N |\beta\nabla_i E(\bm{r}^N)|^2\, \ee^{-\beta E(\bm{r}^N)} \\
=& Z^{-1} \int\dd\bm{r}^N \beta\nabla_i^2 E(\bm{r}^N)\, \ee^{-\beta E(\bm{r}^N)} \\
=& - \langle \beta \nabla_i\cdot \bm{F}_i\rangle_0
\label{eq:F_div_F}
\end{split}
\end{align}
where $Z= \int\dd\bm{r}^N \ee^{-\beta E(\bm{r}^N)}$ is the equilibrium partition function.  The first and third equalities
in (\ref{eq:F_div_F}) follow trivially from the definition of the equilibrium average, while the second relies on an integral by parts.
This result is well known and has been exploited to determine the temperature of a system directly from its configurations~\cite{rugh97,butler98}.
At equilibrium, we conclude that $\langle \Veff \rangle_0 = -\frac\beta4 \sum_i \langle | \bm{F}_i|^2 \rangle_0$.

Data for the two terms in $\Veff$ is shown in Fig.~\ref{fig:FdivF}.  Despite (\ref{eq:F_div_F}),
we note a small difference between the two terms, even at equilibrium.  This effect arises because of
the truncated and shifted Lennard-Jones potential that we use in simulation, which 
has a discontinuity in its first derivative at the cutoff radius $r_{ij}^\mathrm{cut}=2.5\sigma_{ij}$.  

We discuss this effect in Appendix~\ref{app:divF} (see also~\cite{butler98}) 
where we define a regularised average $\langle \nabla\cdot\bm{F}_i \rangle^\mathrm{sim}$,
and discuss how (\ref{eq:F_div_F}) is modified to account for this regularisation.  Consistent with Fig.~\ref{fig:FdivF}, we find that
the effect of this regularisation is small throughout, so we use $\langle \nabla\cdot\bm{F}_i \rangle^\mathrm{sim}$ interchangeably
with $\langle \nabla\cdot\bm{F}_i \rangle$ in what follows.

Having accounted for the small systematic deviation between the two quantities plotted in (\ref{fig:FdivF}),
the most important feature of that figure is that the two contributions to $\Veff$ remain almost equal,
as $s$ increases.  That is, for the range of $s$ considered, our numerical results indicate that
\begin{equation}
\langle \kalt \rangle_{s} \approx -\frac\beta4 \sum_i \langle | \bm{F}_i|^2 \rangle_s
 \approx \frac{1}{4} \sum_i \langle \nabla_i \cdot \bm{F}_i \rangle_s  .
\label{eq:kalt-divF-F2-approx}
\end{equation}
Since (\ref{eq:F_div_F}) applies only at equilibrium, this is a non-trivial result. 
Our interpretation is that the `slow' (structural) degrees of freedom respond strongly to the bias $s$,
while the `fast' (or `vibrational') degrees of freedom respond much more weakly. In other words biasing moves the 
system to a region of the energy landscape not typical of equilibrium, but the system explores that region
as if it were at equilibrium.  If this is indeed the case,
the equilibrium assumption required to prove (\ref{eq:F_div_F}) can be replaced by a weaker, `quasi-equilibrium'
assumption for the fast modes, leading to a similar result.

We formalise this hypothesis within a mean-field description~\cite{ktw89,cavagna09-ped}, assuming that the system has
many metastable states. 
A short relaxation time is associated with intrastate (``vibrational'') motion
and a longer  relaxation time is associated with structural rearrangement (between states)~\cite{jack10-rom}.
We emphasise that metastable states are defined dynamically, by reference to their lifetime~\cite{biroli-kurchan01,jack10-rom}:
each state contains many energy minima (`inherent structures'~\cite{still95}).

Following the discussion of Ref.~\onlinecite{jack10-rom}, for a weak
bias $s$ then the steady state distribution over configurations $\CC$ is
\begin{equation}
p_\mathrm{ss}(\CC) \approx w_{a(\CC)} \ee^{-\beta E(\CC)} / Z_{a(\CC)} 
\label{equ:quasi-eq}
\end{equation}
where $a(\CC)$ is the state containing configuration $\CC$, while $w_{a(\CC)}$ is the probability of that state, and $Z_{a}=\sum_{\CC\in a}\ee^{-\beta E(\CC)}$ is the equilibrium weight of state $a$.  (Here, the short-hand notation $\CC$ indicates a configuration $\bm{r}^N$.)
The $s$-dependence of
(\ref{equ:quasi-eq}) comes only from the weights $w_a$.  If $w_{a}=Z_{a}$ for all states $a$ then we recover the equilibrium Boltzmann distribution (at $s=0$).  
For finite $s$ then one expects the $w_a$ associated with long-lived metastable states to be enhanced.  A similar idea was discussed in Ref.~\onlinecite{jack11-stable}, where
inherent structures were used in place of metastable states.

As usual with mean-field scenarios,
Equ.~(\ref{equ:quasi-eq}) is approximate for (at least) two reasons: firstly, it assumes that each configuration can be assigned to a single metastable state (which neglects configurations on the boundaries between states); secondly it assumes that intra-state fluctuations are unaffected by the field $s$.
The first approximation can be ignored in mean-field models because configurations on boundaries between states have negligible weight in 
$p_\mathrm{ss}(\CC) $.  The second approximation is valid for small $s$, 
if (and only if) fast and slow dynamics take place on well-separated time scales.  This situation is realised
 in mean-field models and may be expressed in terms of a condition on the eigenvalues of the time evolution operator of the system~\cite{jack10-rom}.

In finite-dimensional systems (where mean-field theory is not exact), 
both of these approximations lead to deviations from (\ref{equ:quasi-eq}), 
but one expects that equation to give a reasonable
description of the system if the lifetimes of the metastable (inactive) states are much longer than time scales for motion within these states.
Ref.~\onlinecite{jack11-stable} shows that this condition is quite well-satisfied.
Hence, one may repeat the analysis of Equ.~(\ref{eq:F_div_F}), but using (\ref{equ:quasi-eq}) in place of the Boltzmann distribution.
One arrives at the same conclusion, that the two terms plotted in Fig.~\ref{fig:FdivF} should be equal.  The largest error in that analysis comes
from configurations that lie on boundaries between metastable states~\cite{kurchan96}, but our numerical results indicate that these configurations do not
contribute too much to these averages, and that the quasi-equilibrium hypothesis of (\ref{equ:quasi-eq}) seems to hold quite accurately.
This is the sense in which the slow fluctuations (between states) respond strongly to the field $s$ (via the $w_a$), while the fast
(intra-state) fluctuations respond much more weakly.

\subsection{Vibrational modes of the fluid in biased ensembles}

\begin{figure}
\includegraphics[width=7.5cm]{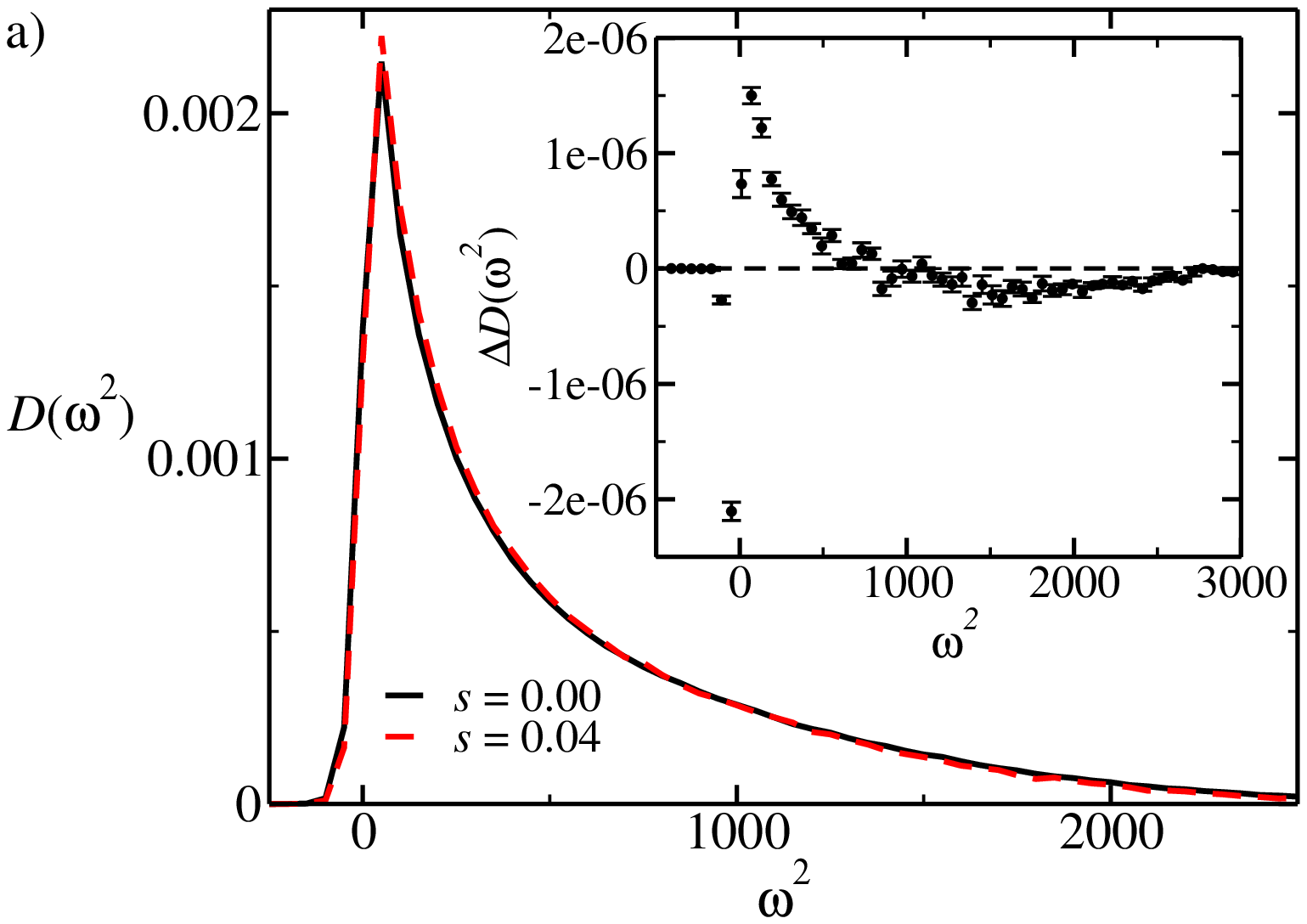}
\includegraphics[width=7.5cm]{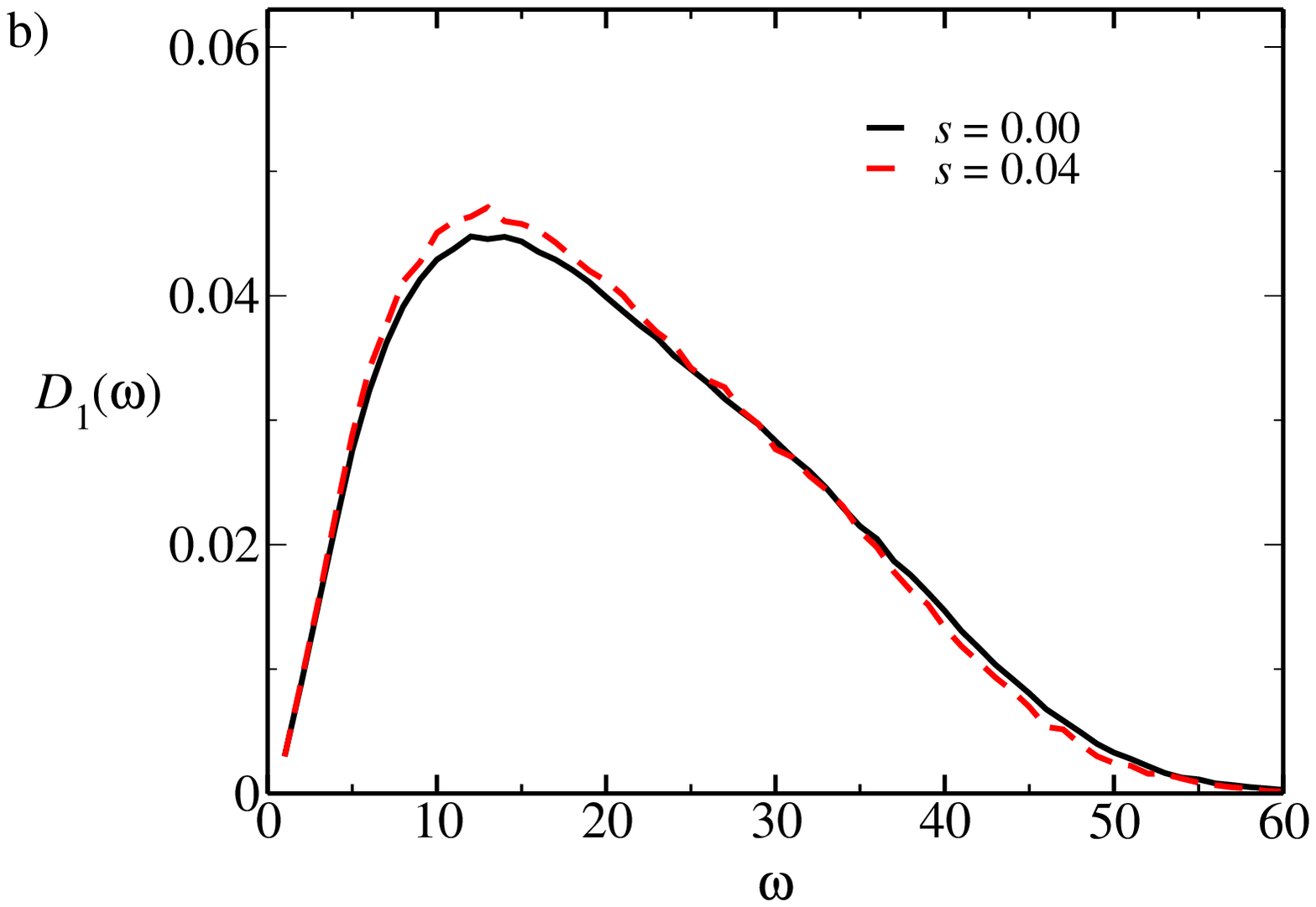}
\caption{(a) Distribution of eigenvalues of the Hessian for both phases.
(a, inset) The difference $\Delta D(\omega^2) = [D(\omega^2)_{s=0.04} - D(\omega^2)_{s = 0.00}]$ 
between the phases. 
The distribution for the active phase is slightly broader, and it associated mean value of $\omega^2$ is larger.
(b) Distribution of $\omega$ where $\omega^2>0$ for both phases.
}\label{fig:modes}
\end{figure}

\begin{figure}
\includegraphics[width=7.5cm]{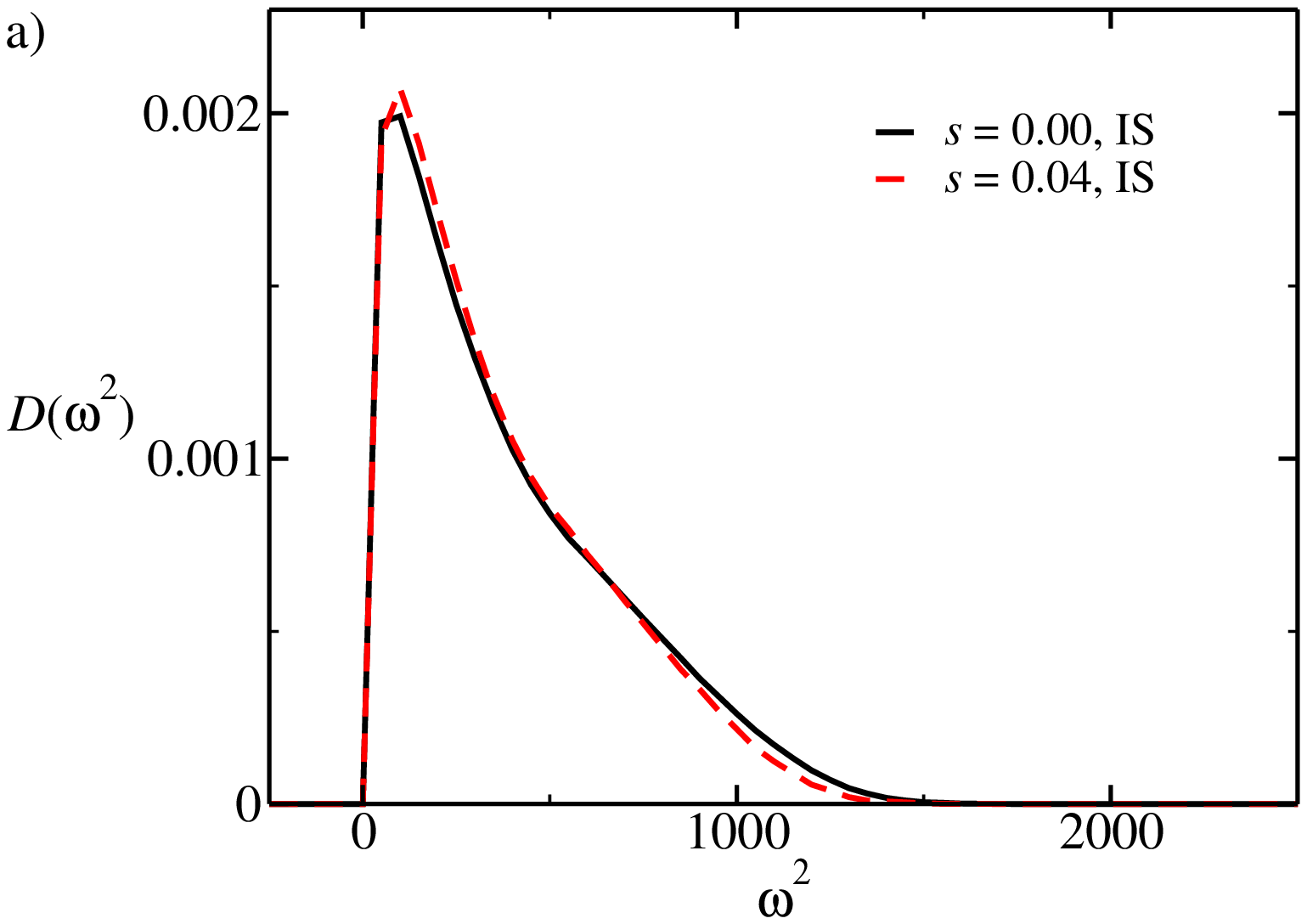}
\includegraphics[width=7.5cm]{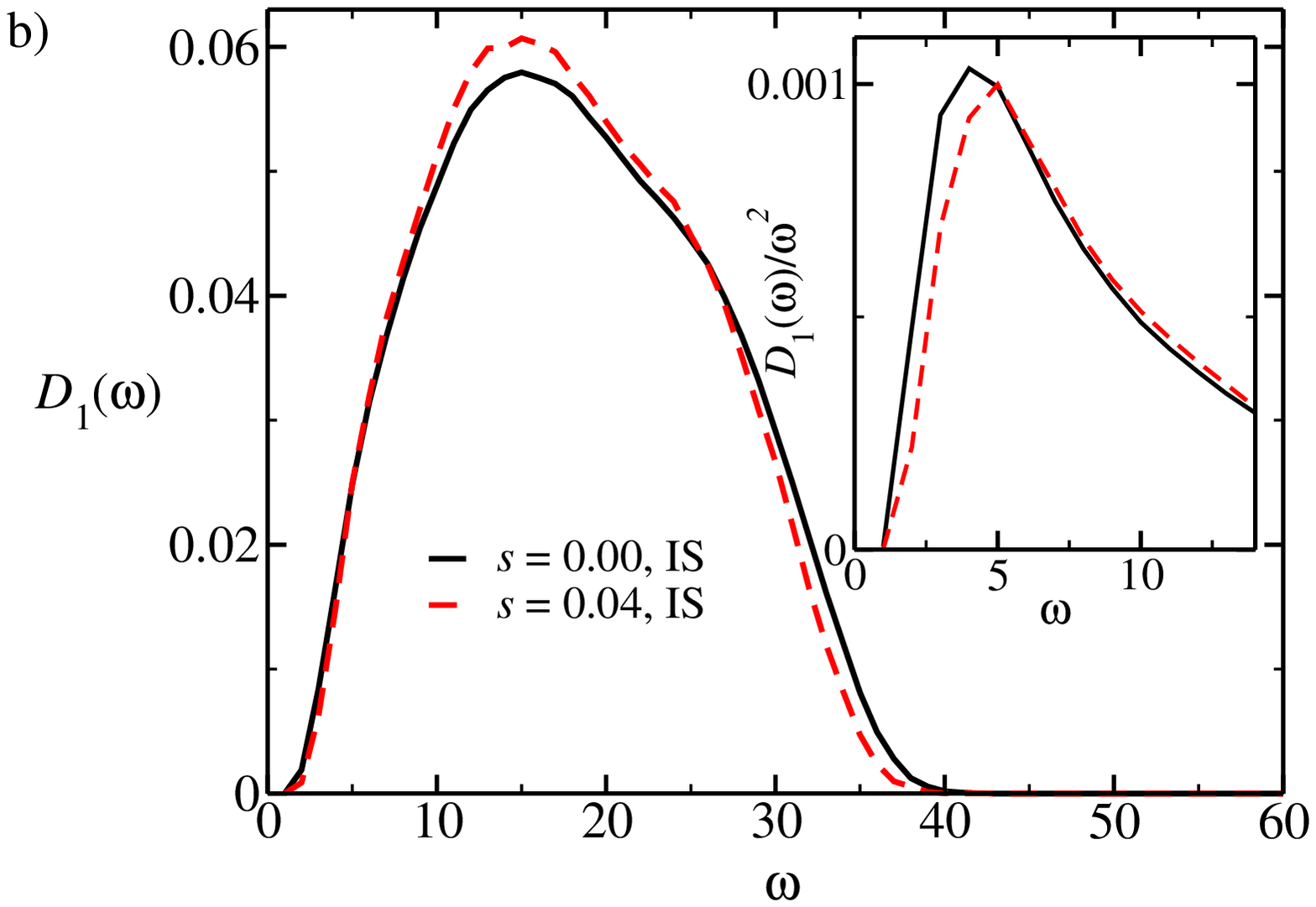}
\caption{(a) Distribution of eigenvalues of the Hessian for inherent structures of both phases.
(b) Distribution of $\omega$ for inherent structures of both phases.
(b, inset) Dividing $D_1(\omega)$ by $\omega^2$ emphasises the lack of low frequency modes associated with the inactive phase.
}\label{fig:modes_IS}
\end{figure}

\begin{figure}
\includegraphics[width=7.5cm]{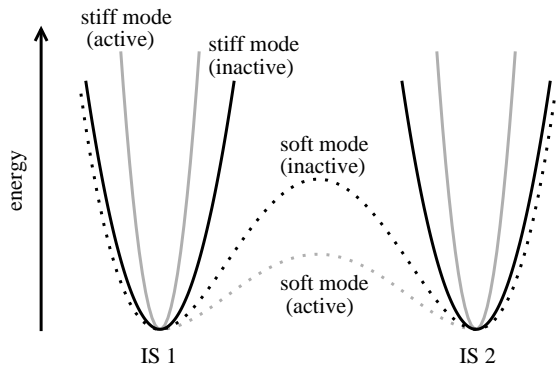}
\caption{A schematic representation of the differences in the energy landscape between the active and inactive phase.
In the inactive phase, the barriers between basins (inherent structures) are smaller making rearrangements on large length scales
 less likely. These correspond to small values of $\omega^2$.
The strongly curving directions around basins are less steep in the inactive phase, allowing more motion on short length scales. These correspond to large values of $\omega^2$.
}\label{fig:landscape}
\end{figure}

The relationship between $\kalt$ and the properties of equilibrium and inactive states can also be 
analysed through the distribution of eigenvalues of the dynamical matrix (or Hessian) $H$.
This distribution, together with the vibrational normal modes of supercooled liquids have been connected with their dynamical properties 
in a variety of studies~\cite{coslovich06,brito07,reichman08-modes,manning11}.   
Here we exploit the connection between the matrix $H$ and the contribution
of $\sum_i\nabla_i\cdot\bm{F}_i$ to $\kalt$.
The Hessian is a $3N \times 3N$ matrix with elements
$
H_{i\mu,j\nu} = \frac{\partial E(\bm{r}^N)}{\partial r^{\mu}_i\partial r^{\nu}_j} ,
$
 where the indices $i$ and $j$ run over all particles and $\mu$ and $\nu$ run over the cartesian components of the position vectors $\bm{r}_i$.

The matrix $H$ has $3N$ eigenvalues, which we denote by $\omega_1^2,\omega_2^2,\dots$.  Here, each $\omega_\alpha$ can be interpreted as a natural
frequency for vibrational motion on the energy landscape, along a particular eigenvector.  
However, we note that since typical configurations of the system are not located at minima of the energy landscape,
some eigenvalues of $H$ will be negative, $\omega_\alpha^2<0$.  In this case the interpretation of $\omega_\alpha$ is less clear,
but the relevant directions on the energy landscape are unstable, indicating that the system is close to a saddle point of the
landscape, and not a stable minimum.
The $\nabla\cdot \bm{F}$ term in $\Veff$ is related to the eigenvalues as
\begin{equation}
\sum_i \nabla_i \cdot \bm{F}_i = -\mathrm{Tr}(H) = -\sum_{\alpha=1}^{3N} \omega_\alpha^2 
\label{eq:divF_H}
\end{equation}

Defining the distribution of eigenvalues, $D(\omega^2)$, the trace can be expressed as
\begin{align}
\langle \mathrm{Tr}(H) \rangle = 3N \int_{-\infty}^{\infty}\!\dd(\omega^2)\, \omega^2 D(\omega^2)
\label{eq:trD}
\end{align} 
Combining (\ref{eq:kalt-divF-F2-approx}) and (\ref{eq:divF_H}) and (\ref{eq:trD}), we see that $\kalt\approx  \frac{-3N}{4}\int_{-\infty}^{\infty}\!\dd(\omega^2)\, \omega^2 D(\omega^2)$,
allowing us to relate the difference in $\kalt$ between active and inactive (small-$k$) states
to the distribution $D(\omega^2)$ of these states.  Results are shown in Fig.~\ref{fig:modes}.  Comparing equilibrium ($s=0$) and inactive ($s>0$)
data, the differences in $D(\omega^2)$ are subtle,
but the dominant effect is that the main peak in $D(\omega^2)$ is slightly sharper in the inactive state.  That is, the inactive
state has fewer modes with small or negative $\omega^2$, but also fewer modes with large positive $\omega^2$.  Hence it has more modes
with intermediate $\omega^2$.  When evaluating the change in $\mathrm{Tr}(H)$ between states, the dominant effect comes from large eigenvalues,
which correspond to ``stiff'' (strongly-curving) directions on the potential energy landscape.  Fig.~\ref{fig:modes} shows 
that there are fewer stiff directions in the inactive
state, and this results in $\kalt$ being larger (less negative) for that state. The difference is more pronounced when plotting
$D_1(\omega)$, the distribution of $\omega$ among modes where $\omega^2 > 0$.

In Fig.~\ref{fig:modes_IS}, we show the distributions of $\omega^2$ and of $\omega$ that we obtained by using conjugate gradient
minimisation on configurations from the $s$-ensemble, and then constructing the matrix $H$ at the resulting energy minimum [inherent structure (IS)].  In this case, all eigenvalues of $H$ are positive.  The differences in $D(\omega^2)$ 
between active and inactive
states are more pronounced at the IS level, but the main conclusion is the same: the peak in $D(\omega^2)$ is narrower in the inactive
state, and this pushes the mean value of $\omega^2$ to a smaller value.  However, these data also emphasise that the inactive state
has fewer ``soft'' modes (with small $\omega)$, compared to equilibrium.  This effect was noted in Ref.~\onlinecite{jack11-stable}: it indicates that part of the 
stability of the inactive state can be accounted for by the paucity of soft-directions on the energy landscape.

The resulting physical picture is summarised in Fig.~\ref{fig:landscape}.  The potential energy surface (or `landscape') is divided into basins,
each associated with a single inherent structure (local minimum).  Moving away from the inherent structure, most of the directions 
are quite `stiff', with large $\omega$, but a few are `soft', with small $\omega$.  Comparing the equilibrium state with the inactive (small-$k)$ state,
Figs.~\ref{fig:modes} and~\ref{fig:modes_IS} show that the stiff directions in the inactive state are (on average) less stiff than at equilibrium;
on the other hand, the soft directions in the inactive state are also less soft than at equilibrium.  The activity parameter $\kalt$ of Pitard~\etal~\cite{pitard11}
is most sensitive to the stiff directions: the stiffer these are, the less particles are free to move (on short scales), and the smaller is $\kalt$.
On the other hand, the activity parameter $k$ of Hedges~\etal~\cite{hedges09} is most sensitive to structural relaxation, which couples more strongly to the soft modes: 
these are less soft in the inactive state, suppressing large-scale particle motion, and reducing $k$.

This difference in sensitivity to fast and slow motion explains the anticorrelation between $k$ and $\kalt$ in Fig.~\ref{fig:scatter}, and it also
explains why the active/inactive transition of Ref.~\onlinecite{hedges09} appears only in $\salt$-ensembles with $\salt<0$.  We argue that it should be borne in mind in any future studies that use
$\Veff$ to measure activity.

\subsection{Liquid structure in biased ensembles}

We now turn to the structure of the active and inactive states that we have found, and the connection of this structure to $\kalt$.
It is notable from Fig.~\ref{fig:K_alt_K_s} that typical values of $\kalt$ are around $-380(\epsilon/\sigma^2)$, while
the difference in $\kalt$ between active and inactive states is much smaller, around $30(\epsilon/\sigma^2)$.  (We give the units
of $\kalt$ explicitly in this discussion: recall that numerical data are shown after fixing $(\epsilon,\sigma)$ to unity.)  

To interpret these results, it is useful to write
\begin{equation}
\big\langle \sum_{i}\nabla\cdot\bm{F}_i \big\rangle_s = \sum_{i\neq j} \int 4\pi r^2 \mathrm{d}r\, 
\tilde{g}_{ij}(r) \nabla^2 V_{ij}(r)  
\label{eq:FG}
\end{equation}
where $\tilde{g}_{ij}(r) = \langle \sum_{j\neq i}\delta(r-r_{ij}) \rangle_s$ is proportional to a
radial distribution function (in the $s$-ensemble).  
Since $V_{ij}(r)$ and $\tilde{g}_{ij}(r)$ depend on the particle indices $i$ and $j$ only through their types, it is convenient to
use a shorthand notation for the non-trivial part of the integrand in (\ref{eq:FG})
\begin{equation}
G^{\rm AA}(r) = \left.\nabla_i^2 V_{ij}(r) \tilde{g}_{ij}(r)\right|_{i,j\, {\rm of\, type\, A}}
\end{equation}
where the right hand side is evaluated with $i$ and $j$ both being particles of type A. 
Similarly, we define $G^{\rm AB}(r)$ and $G^{\rm BB}(r)$ for particles of other types.  
(Note that these functions depend implicitly on the biasing parameter $s$, through $\tilde{g}_{ij}$.) 

By comparing $4\pi r^2 G^{\mathrm{AA}}(r)$ to $g_{\mathrm{AA}}(r)$ (the radial distribution function 
for particles of species A), we can see how the liquid structure on different length scales 
contributes to $\langle \nabla.\bm{F}_i \rangle_s$.  
We focus only on the function for the large particles as these are the most numerous species.

\begin{figure}
\includegraphics[width=7.5cm]{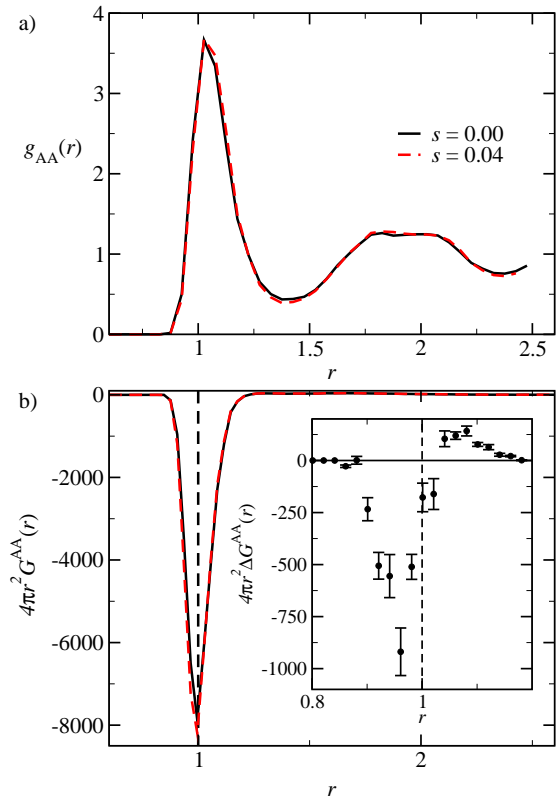}
\caption{(a) Comparison of the partial pair correlation function for large particles between the active and inactive phase. Although there are some differences (the height of the first peak and the depth of the first trough) they are small.
(b) The function $4\pi r^2 G^{\mathrm{AA}}(r)$ which can be integrated to give $\langle \nabla \cdot \bm{F}_i \rangle_s$. The interesting part of this function occurs around the position of the first peak in the pair correlation function. The inset panel shows the difference in this function between the phases, $\Delta G^{\mathrm{AA}}(r) = [G^{\mathrm{AA}}(r)]_{s=0.04}-[G^{\mathrm{AA}}(r)]_{s=0.00}$. This serves to illustrate that the changes in $\Kalt$ come from structural changes on short length scales.
}\label{fig:gr_vr_r2}
\end{figure}

Fig. \ref{fig:gr_vr_r2} (a) shows $g_{\mathrm{AA}}(r)$ for the active phase (at $s = 0.00$) and the 
inactive phase (at $s = 0.04$). There are some subtle changes: the first and second peaks and the first trough
are enhanced in the inactive phase. Panel (b) shows $4\pi r^2 G^{\mathrm{AA}}(r)$ for the same values of $s$.
Only a small region contributes to $\langle \nabla.\bm{F}_i \rangle$ - the width is less than that of 
the first peak in $g_{AA}(r)$. This further emphasises that $\kalt$ is dominated by behaviour on short length scales.
Again, the differences between the phases are subtle. This is in line
with the observation that the size of the change in $\kalt$ between phases is much smaller than the size of $\kalt$ itself.

To emphasise the change, we consider the difference $\Delta G^{\mathrm{AA}}(r) = [G^{\mathrm{AA}}(r)]_{s = 0.04} - [G^{\mathrm{AA}}(r)]_{s=0.00}$. 
This is shown in the inset to figure \ref{fig:gr_vr_r2} (b).
It is clear that the change in $\kalt$ is largely due to changes in the liquid structure at very small length scales; the dashed line
in the plot indicates where $G^{\mathrm{AA}}(r)$ is largest in magnitude, which corresponds to the maximum of the
first peak in $g_{\mathrm{AA}}(r)$. These changes are subtle enough that they are not apparent when comparing radial
distribution functions, but since $\nabla^2 V_{ij}(r)$ is very large for small $r$ they are ultimately what is important
when considering $\kalt$.

In addition to the results in Fig.~\ref{fig:gr_vr_r2}, 
we have obtained similar data for $G^{AB}(r)$ and $G^{BB}(r)$: the main picture is the same but the smaller
numbers of B particles in the system mean that these functions contribute less strongly to $\Veff$, and also that the numerical
uncertainties in our results are larger.  As shown by Speck and coworkers~\cite{speck12-jcp,speck12-prl}, 
 the radial distribution function
$g^\mathrm{BB}(r)$ shows the largest relative
changes between active and inactive states.  However, the small number of B-particles means that this gives a relatively
small contribution to the changes in $\kalt$ shown in Fig.~\ref{fig:K_alt_K_s}.

\subsection{The dynamical action}

Finally, we discuss one other context in which the activity $\Kalt$ appears.
For overdamped dynamics as in (\ref{eq:langevin}), at equilibrium,
the probability of a trajectory $\rnt$ can be written as~\cite{autieri09}
\begin{align}
P_0[\rnt] &= \frac{1}{\cal Z} P_\mathrm{free}[\rnt]\cdot \mathrm{e}^{\frac{\beta}{2}[E(0) - E(\tobs)]} \nonumber \\ & \qquad\qquad \times\mathrm{e}^{-\beta D_0 \Kalt[\rnt]}
\end{align}
where $P_{\rm free}[\rnt]$ is the probability of the trajectory in the absence of any forces, and $\cal Z$ is a normalisation constant.

Hence if we consider the equilibrium distribution of $\kalt$ for this model, we have
\begin{align}
P_{s=0}(\kalt) = \frac{1}{\cal Z} \ee^{N\tobs[{\cal S}(\kalt) - \beta D_0\kalt]},
\end{align}
where $\ee^{N\tobs{\cal S}(\kalt)}$ is the marginal distribution of $\kalt$ associated with the distribution 
$P_\mathrm{free}[\rnt]\mathrm{e}^{\frac{\beta}{2}[E(0) - E(\tobs)]}$.  (We emphasise
that the function ${\cal S}(\kalt)$ depends on the parameter $\beta$ via the definition of $\kalt$, and it also depends on $D_0$.)
Further,
the distribution of $\kalt$ within the $\salt$-ensemble is 
\begin{align}
P_{s}(\kalt) \propto \ee^{N\tobs[{\cal S}(\kalt) - (\beta D_0+\salt)\kalt]},
\label{eq:ps-kalt}
\end{align}

There is a relevant analogy here: compare the distribution of the energy density $e=E/N$ in a thermal system at equilibrium,
\begin{align}
P_{\beta}(e) \propto \ee^{N[S(e) - \beta e]},
\label{equ:gibbs}
\end{align}
where $S(e)$ is the entropy per particle.  This analogy between ensembles of trajectories like (\ref{eq:ps-kalt}) and ensembles
of configurations like (\ref{eq:ps-kalt}) was a key starting point for studies of the dynamical transitions and biased ensembles that
we consider here~\cite{merolle05,jack06-spacetime,lecomte07,garrahan-fred09}.

Extending this analogy, the interpretation of $\kalt$ and $\salt$ is as follows.  Within the distribution $P_0[\rnt]$, there are
many trajectories with large values of $\kalt$, each of which is individually rare because of the factor of $\ee^{-\beta D_0 \Kalt}$.
There are fewer trajectories with smaller $\kalt$, but these are individually more probable because they are less strongly
suppressed by the factor $\ee^{-\beta D_0 \kalt}$.  The most likely value of $\kalt$ occurs when the `entropic' term ${\cal S}(\kalt)$
balances the `energetic' term $\beta D_0 \kalt$.  [Here we are using the labels `entropic'/`energetic' to emphasise the
 analogy with (\ref{equ:gibbs}): these terms have no simple relation to thermodynamic energy or entropy.]
 
If we introduce a negative value of $\salt$, the system is biased towards the more numerous (`entropically favourable') trajectories
in the system, which have larger (or less negative) values of $\kalt$.  As shown in Fig.~\ref{fig:K_alt_K_s}, even a small negative $\salt$
is sufficient to
drive the system into an `inactive' state in which structural relaxation is arrested.
The unexpected anticorrelation between $k$ and $\kalt$ that we found in this study arises because the inactive state has the higher
`entropy' $\cal S$ in trajectory space.  The reason for this is that the inactive state consists of configurations in   
in which most directions on the energy landscape are not too `stiff': despite the slow structural relaxation, 
the particles have greater freedom to move on small length scales, compared with equilibrium.  
And the more free the particles are to move, the more trajectories are available, and the larger is $\cal S$.
As before, the conclusion is that propensity for motion on small scales is anti-correlated with propensity on scales of the order
of the particle diameter.

\section{Conclusions and outlook}
\label{sec:conc}

This study has two central conclusions.  Firstly, the transition found by Hedges~\etal~\cite{hedges09} for $s>0$ corresponds
to a transition for $\salt<0$ within the ensembles defined by Pitard~\etal~\cite{pitard11}.  
Secondly, the activity parameter $\kalt$ defined in Ref.~\onlinecite{pitard11}
couples to dynamical motion on small scales, which is anticorrelated with the structural relaxation of the fluid.  This
anticorrelation arises from properties of the energy landscape of the inactive state.  In addition to these main points, we
have also discussed the structure of the inactive states and the connection of $\kalt$ the liquid structure; and also the
extent to which the inactive states have the quasi-equilibrium property given in (\ref{equ:quasi-eq}).

We hope that this work clarifies the role of the activity measurement introduced by Pitard~\etal~\cite{pitard11},
which we have denoted by $\Kalt$.  Equ.~(\ref{eq:ps-kalt}) shows that $\Kalt$
is intimately connected with dynamical motion in overdamped Langevin systems, and it is also strongly connected
to the energy landscape of the fluid.  These facts present a strong argument in favour of $\Kalt$ as an activity
measure that arises naturally from the dynamics of the system, without any prejudice as to the nature of its dynamical
relaxation.  However, the results of Fig.~\ref{fig:scatter} show that $\Kalt$ must be interpreted carefully,
since the extent of short-scale motion may not be correlated with the effectiveness of structural relaxation.
Also, this study did not find
evidence for singular behaviour in $\langle\kalt\rangle_{\salt}$ for the range of positive $\salt$ that we considered: 
 the physical interpretation of the
behaviour found in Ref.~\onlinecite{pitard11} for larger positive $\salt$ remains unexplained (although it seems unrelated
to the active/inactive crossover discussed in Ref.~\onlinecite{hedges09}).

\begin{acknowledgments}
We thank Fred van Wijland, Vivien Lecomte and Estelle Pitard for helpful discussions.  
We are grateful to the EPSRC for support through grant EP/I003797/1.
\end{acknowledgments}

\begin{appendix}
\section{Sampling biased ensembles}
\label{app:sampling}

We sample trajectories from the $s$-ensemble and $\salt$-ensemble by using
transition path sampling (TPS).
This method samples trajectories in a similar way to the sampling of configurations by standard Metropolis Monte Carlo methods.
Its operation is reviewed in Ref.~\onlinecite{tps-annrev02} and the `shifting moves' used in this study are discussed in Ref.~\onlinecite{dellago98}.  We give a brief overview
here:
Starting with an initial trajectory $\bm{r}_0^N(t)$,
a new trajectory $\bm{r}_1^N(t)$ is generated by a `shifting move'.  
In `forward shifting', one chooses a random number $p$ between $1$ and $M$, and slices $1,2,\dots, p$
of $\bm{r}_0^N(t)$ are discarded.  The remaining slices ($p+1,\dots, M$) of $\bm{r}_0^N(t)$ form the initial slices ($1,\dots, M-p$) of
the new trajectory $\bm{r}_1^N(t)$.  Slices $M-p+1,\dots, M$ are then generated by unbiased dynamical evolution from
slice $M-p$.  
Finally, this new trajectory $\bm{r}_1^N(t)$ is accepted with probability
\begin{align}
P_{\mathrm{acc}} = \mathrm{min}\left\{1, \ee^{ - sK[\bm{r}_1^N\!(t)] + sK[\bm{r}_0^N\!(t)]} \right\}.
\end{align}
Otherwise one rejects the new trajectory and retains the original one, $\bm{r}_0^N(t)$.
This procedure is used in conjunction with ``backwards shifting'' moves where slices $1,2,\dots,p$ of $\bm{r}_0^N(t)$ are used
to form slices $M-p+1,\dots, M$ of $\bm{r}_1^N(t)$, and then slices $1,\dots,M-p$ of $\bm{r}_1^N(t)$ are generated by unbiased time evolution,
backwards in time from slice $M-p+1$ (use of this scheme requires the time-reversal symmetry property of the equilibrium state of
the model).  This combination of moves ensures detailed balance within the ensemble of trajectories (\ref{eq:traj_prob}),
so after sufficiently many moves, the procedure converges in a stationary regime which generates representative samples
of the ensemble.
Further, since the system is stochastic and the ensemble of trajectories being sampled is (approximately) time-translationally
invariant, these shifting moves are effective in sampling the ensemble, and it is not necessary to supplement them with
`shooting' moves. (A combination of shooting and shifting is the conventional choice in rare event sampling problems dominated
by barrier crossing, but we do not use this procedure here).

The results shown here were obtained from TPS simulations as follows.  We used a weighted histogram analysis (WHAM)~\cite{wham89} to 
combine data obtained using different values of $s$ and $\salt$.
For trajectories of length $\tobs = 200\Delta t$ we used data from $s=-0.025$ to $s=0.03$ in the $s$-ensemble 
and from $\salt= -3.0\times 10^{-5}$ to $\salt= 5.0\times 10^{-5}$ in the $\salt$-ensemble.
For trajectories of length $\tobs = 400\Delta t$ we used data from $s=0.00$ to $s=0.020$ for the $s$-ensemble 
and from $\salt = -1.75\times 10^{-5}$ to $\salt= 0.00$ for the $\salt$-ensemble.
These choices ensure that we concentrate our numerical effort in the crossover regime between active and inactive states: 
as we bias further into the inactive regime, the slow structural dynamics of
the inactive state limit the effectiveness of sampling.  We therefore access the inactive regime by histogram reweighting
from the crossover regime, using the results from WHAM.

Large values of $s$ (and $-\salt$) bias the system towards inactive states, and this can lead to crystallisation within trajectories.
This happens rarely and we exclude trajectories with a high degree of crystalline order from our analysis.
We measure crystalline order using the common neighbour analysis scheme described in the supplement to Ref. \onlinecite{hedges09}.
We note that the values given for the maximum separation of bonded pairs of particles in Ref. \onlinecite{hedges09} are incorrect, and we use the correct
values: $\lambda_{\rm{AA}} = 1.45$, $\lambda_{\rm{AB}} = 1.25$ and $\lambda_{\rm{BB}} = 1.07$.

We note that Pitard~\etal~\cite{pitard11} used a different method~\cite{giardina06} to sample biased ensembles of trajectories.  In contrast
to transition path sampling, which operates on trajectories of fixed duration $\tobs$, that method provides direct estimates
of observables in the limit where $\tobs\to\infty$. On the other hand, the algorithm requires that many copies (or clones) of
the system evolve in parallel, and there are systematic errors associated with the method~\cite{giardina06}, which
vanish only when the number of clones is taken to infinity.  In this sense, the TPS method results in controlled sampling
of ensembles with finite $\tobs$, requiring an extrapolation to reach the large-$\tobs$ limit; on the other hand,
the method of Ref.~\cite{giardina06} gives direct access to a limit of large $\tobs$, 
but at the expense of an extrapolation in the number of clones. 

\section{Regularisation of $\nabla\cdot\bm{F}_i$}
\label{app:divF}

The results in Fig.~\ref{fig:FdivF} indicate that Eq.~(\ref{eq:F_div_F}) is not satisfied exactly at equilibrium,
for the model system used here.  As discussed in Ref.~\onlinecite{butler98}, this behaviour is generic for systems
where interaction potentials are truncated.  
To analyse this behaviour quantitatively, we imagine modifying the potential 
$V_{ij}(r_{ij})$ in a region of width $\varepsilon$ around $r_{ij}^\mathrm{cut}$
so that its second derivative exists everywhere, and then taking the limit of small $\varepsilon$.
In this case,
\begin{align}
\nabla_i\cdot\bm{F}_i = \sum_{j(\neq i)}\left[q_{ij} + \tilde{q}_{ij}\delta(r_{ij} - r_{ij}^{\mathrm{cut}})\right]
\label{eq:F_divF_correct}
\end{align}
where
\begin{align}
q_{ij} = \left\{ \begin{array}{ll}  -\nabla^2 V_{ij}(r_{ij}), & r_{ij}<r_{ij}^{\mathrm{cut}} \\
     0 & \mbox{otherwise}, \end{array} \right. 
\end{align}
and $\tilde{q}_{ij} = \frac{\mathrm{d}V_{ij}(r_{ij}^\mathrm{cut})}{\mathrm{d}r_{ij}}$ is the discontinuity in the force at
the potential cutoff.  If one  uses (\ref{eq:F_divF_correct}) as the \emph{definition}
of $\nabla\cdot\bm{F}_i$, then (\ref{eq:F_div_F}) will hold exactly at equilibrium.

However, the $\delta$-function in (\ref{eq:F_divF_correct}) makes it problematic in simulation.
We therefore define instead
\begin{equation}
\big\langle \sum_i \nabla\cdot\bm{F}_i \big\rangle^\mathrm{sim} = \big\langle \sum_{i\neq j} q_{ij} \big\rangle
\end{equation}
and note that
\begin{align}
\big\langle \sum_i \nabla\cdot\bm{F}_i \big\rangle = & \big\langle \sum_i \nabla\cdot\bm{F}_i \big\rangle^\mathrm{sim} 
+ N_\mathrm{A} \rho_\mathrm{A} \Delta_{\rm AA} 
\nonumber \\ & \qquad 
+ N_\mathrm{A} \rho_\mathrm{B} \Delta_{\rm AB} 
+ N_\mathrm{A} \rho_\mathrm{A} \Delta_{\rm BB} 
\label{eq:divF-correction}
\end{align}
where $\langle \sum_i \nabla\cdot\bm{F}_i \rangle$ on the left hand side uses the definition
from (\ref{eq:F_divF_correct}), while $\Delta_{\rm AA} = 4\pi (r_\mathrm{AA}^\mathrm{cut})^2 \tilde{q}^{\rm AA} g^\mathrm{AA}(r_\mathrm{AA}^\mathrm{cut})$,with
similar expressions for $\Delta_{\rm AB},\Delta_{\rm BB}$.  
Here $\rho_{\rm A}=N_\mathrm{A}/V$ is the number density of A-particles, 
$g^\mathrm{AA}(r)$ is the radial distribution function between A particles, and $\tilde{q}^\mathrm{AA}$
is the value of $\tilde{q}_{ij}$ if particles $i,j$ are both of type A.  (We used the fact that if particles
$i$ and $j$ are of type A then $\langle \delta(r-r_{ij})\rangle = 4\pi r^2 \rho_{\rm A} g^{\rm AA}(r)$).
We have evaluated the $\Delta$-terms in (\ref{eq:divF-correction}) at equilibrium, and verified that the data
in Fig.~\ref{fig:FdivF} are then consistent with~(\ref{eq:F_div_F}).  However, since these $\Delta$-terms are small,
we use $\langle \sum_i \nabla\cdot\bm{F}_i \rangle^\mathrm{sim}$ throughout this work as our numerical estimator
for $\langle \sum_i \nabla\cdot\bm{F}_i \rangle$.

\end{appendix}

\bibliography{glass-rlj.bib}

\end{document}